\documentstyle[preprint,eqsecnum,aps]{revtex}
\begin{document}
\draft
\title{Wulff Construction for Deformable Media}
\author{Joseph Rudnick and Robijn  Bruinsma}
\address{Department of Physics, University of California, Los Angeles,
California 90024}
\date{\today}
\maketitle
\begin{abstract}
A domain in a Langmuir monolayer can be expected to have a shape that
reflects the textural anisotropy of the material it contains. This
paper explores the consequences of $XY$-like ordering. It is found
that an extension of the Wulff construction allows for the calculation
of two-dimensional domain shapes when each segment of the perimeter
has an energy that depends both on its orientation and its location.
This generalized Wulff construction and newly-derived exact
expressions for the order parameter texture in a circular domain lead
to results for the shape of large domains. The most striking result is
that, under general conditions, such domains will inevitably develop a
cusp. We show that the development of cusps is mathematically related
to phase transitions. The present approach is equivalent to a Landau
mean-field version of the theory.

\end{abstract}


\section{Introduction} \label{I}

The regularities in the shapes of crystalline materials have long been
understood to be macroscopic expressions of the positional order of
crystals at the atomic level. The angles between the faces of a
crystal are, for instance, structural invariants dependent only on a
certain set of integers characterizing the faces (Miller indices
\cite{Miller}). In a classic paper \cite{Wulff}, Wulff developed a
geometrical construction allowing for the determination of the shape
of a crystal, provided one knows the values of the surface energy of
the crystal as one varies the Miller indices. One ``draws'' planes
with every possible set of Miller indices. The distance of each plane
from a fixed point in space is proportional to the energy per unit
area of a surface parallel to that plane. The inner envelope of the
planes is then the equilibrium shape of a finite piece of crystal.

Key to the Wulff construction is the dependence of the interfacial
energy on orientation. Anisotropic surface energies are, however, not
restricted to the crystalline solids. A portion of the surface of a
liquid-crystalline mesophase has, in general, a surface energy whose
value depends on its orientation relative to the optical axis. The
principal difference between this system and a crystal is that liquid
crystals may have crystallographic directions that are easily
deformable (e.g. smectics) or they may lack crystallographic order
altogether---and only exhibit orientational order (e.g. nematics). It
seems obvious that the internal softness of liquid crystals and like
systems will suppress such characteristic surface features of solid
crystals as facets and edges. In the case of an extremely soft system
the optical axis ought to adjust itself so as to minimize the
effective surface energy for all orientations. A sample is then
expected to relax to a spherical shape to minimize its overall
interfacial energy. According to this argument, one should expect to
encounter edge-like features only in liquid-crystalline materials that
are relatively rigid. As we will see in this paper, this last
conclusion is, in fact, quite incorrect. Edges can be seen even in
very soft liquid crystals.

The possibility of sharp edges, or cusps, in liquid crystalline
materials was first suggested by Herring in the early 1950's. He
utilized the Wulff construction to demonstrate that liquid crystals
with rigid orientational order can have interfaces with sharp cusps,
provided that the  surface energy is sufficiently anisotropic. The
resulting cusp angle are no longer geometrical invariants but instead
depend on material parameters. It was also noted by Burton, Cabrera
and Frank (BCF) in 1951 that the Wulff construction acquires important
simplifications in two dimensions ($D=2$)\cite{Burtonetal}. Its
derivation becomes a simple exercise in the calculus of variations, in
contrast to the situation in three dimensions, for which there is no
entirely rigorous and general demonstration of the correctness of the
Wulff construction\cite{Herring}. To implement the Wulff construction
in two dimensions, one solves a simple linear differential equation,
while in three dimensions the only known method is the geometrical
construction described above.

Anisotropic shapes of samples in thermal equilibrium are, indeed,
encountered in experiments on $D=2$ mesophases. The systems in
question are Langmuir monolayers consisting of surfactant molecules at
an air-water interface \cite{KnoblerandDesai}. The phase diagram of
such materials generally contain, in addition to the $D=2$ analogue of
the solid liquid and gaseous phases, the so-called liquid condensed
(LC) phase, which is liquid-crystalline. The LC phases are two
dimensional anisotropic liquids in which the direction of the tilted
hydrophobic tails define an anisotropy axis, $\hat{c}$. Because of the
anisotropy between the polar head groups and the non-polar tail there
is no $\hat{c}\ \rightarrow -\hat{c}$ symmetry (as is the case for
nematic liquid crystals).

At the phase boundary between the LC and isotropic liquid phases,
stable coexistence droplets are observed with shapes that are, in
general, non-circular. Recently, polarized fluorescence microscopy
studies of coexistence droplets of pentadecanoic acid with a linear
dimension in excess of 25 $\mu$m have reported that the boundaries of
these droplets contain a {\em cusp} \cite{KnoblerandSchwartz}. Such a
cusp in illustrated in Fig. 1a. The interior angle of the cusp was
found to increase with the size of the domain.

The existence of cusps on the boundary of these droplets would not
seem to be too surprising, as droplets in surfactant systems have been
found to contain toplogical defects in the pattern---or texture---of
the anisotropy axis. A defect located at the boundary could well
produce a cusp. However, Brewster angle microscopy studies---which
allow for detailed visualization of textures---on these materials have
revealed defect-free texures of the ``virtual boojum'' type (see Fig.
1b)\cite{MeunierandRivier}, with the cusp in the sample shape located
opposite the focus of the boojum.

The determination of the shape of these systems demands a simultaneous
minimization of the free energy with respect to the {\em shape} and
the {\em internal structure} of the sample. In other words, we must
find the analogue of the Wulff construction for deformable media. The
question of finding such a generalized Wulff construction for Langmuir
layers belongs to a wider class of shape minimization problems, as
encountered, for instance, in ferrofluid droplets \cite{LangSeth},
where the shape energy depends on the internal structure of the sample
far from the surface. Analytical treatment of such problems beyond
perturbation theory has proven difficult because the coupling between
different parts of the surface, as mediated by the deformable bulk
structure, is highly non-local. As suggested by BCF \cite{Burtonetal},
shape calculations are more easily carried out in $D=2$. We will
demonstrate in this article that a generalized Wulff construction can
be found in $D=2$ for the case when the internal structure of the
sample is describable by an $X-Y$ model. As discussed below, Langmuir
monolayers in the LC have an order parameter which, under certain
conditions, reduces to an $X-Y$ model. We will show that cusp
singularities ought to be generic features of $2D$ drops with an $X-Y$
order parameter.

We will start in Section \ref{III} with a discussion of the free
energy minimization for a fixed sample shape (a circle) and the
resulting textures. In Section \ref{III} we will keep the texture
fixed and minimize the free energy with respect to sampl shape,
following the method of BCF \cite{Burtonetal}. Simultaneous
minimization of the free energy with respect to both shape and sample
texture can be done perturbatively, as discussed in Section \ref{V}.
Our general method, based on complex function theory, is discussed in
Section \ref{VII}. A brief summary of the method has been published
elsewhere \cite{R&B}.

\section{Free energy of XY droplets} \label{III}

In this Section we define the Hamiltonian and examine the order
parameter textures. We will assume a rigid circular boundary, so free
energy minimization simply involves finding the lowest energy texture
consistent with the boundary energy. The order parameter is taken to
be the unit vector $\hat{c}=(\cos \Theta,\sin \Theta)$ with an
associated effective Hamiltonian
\begin{equation}
{\rm H}\left[\Theta(x,y)\right]=\int_{{\rm
interior}}\frac{\kappa}{2}\left|\vec{\nabla} \Theta
\right|^2\,dx\,dy+\int_{{\rm boundary}} \sigma\left(\theta-\Theta
\right)\,dS. \label{1}
\end{equation}
The first term in Eq. (\ref{1}) describes the free energy cost
associated wih a posistion dependence of the order parameter
$\hat{c}({\bf r})$. The coefficient $\kappa$ is, for the $XY$ model,
the so-called spin-wave stiffness. For Langmuir monolayers in the $LC$
phase, $\kappa$ corresponds to the Frank constant for the special case
that the so-called ``splay'' and ``bend'' Frank constants are equal.
The second term on the right hand side or Eq. (\ref{1}) represents the
``surface'' energy of the domain. The variable $\theta$ is the angle
that the unit normal makes with respect to the axis to which the
director angle is referred.

 If the boundary has a fixed shape, minimization of the energy in Eq.
(\ref{1}) implies the following two extremum equations
\begin{equation}
\nabla^2 \Theta(x,y)=0,  \label{2}
\end{equation}
and
\begin{equation}
\kappa \frac{\partial \Theta(x,y)}{\partial n} -
\sigma'(\theta - \Theta) =0. \label{3}
\end{equation}
Eq. (\ref{2}) applies in the bulk, and Eq. (\ref{3}) at the surface of
the domain. The derivative $\partial \Theta/\partial n$ in Eq.
(\ref{3}) is along the surface normal, and
\mbox{$\sigma'(x)=d\sigma(x)/dx$}. The bulk extremum equation, Eq.
(\ref{2}), requires a $\Theta(x,y)$ that is a harmonic function. The
general solution of Eq. (\ref{2}) in $D=2$ can
then be written, for the case at hand, as:
\begin{equation}
\Theta(x,y)=\frac{1}{i}\left(f(x+iy)-f(x-iy)\right), \label{4}
\end{equation}
with $f(z)$ an arbitrary analytic function of $z=x+iy$.
As an example of this method, expand $f(z)$ in a Taylor series:
\begin{equation}
f(z)=\sum_{n=1}^{\infty}\frac{\alpha_n}{2n}z^n  \label{5}
\end{equation}
Then, using Eq. (\ref{4}), we find
\begin{mathletters} \label{6}
\begin{eqnarray}
\Theta(x,y)&=&\alpha_1y+\alpha_2yx+\alpha_3y(3x^2-y^2)+ \cdots
\label{6a} \\
&=&\frac{1}{i}\left[\sum_n\frac{\alpha_n}{2n}(x+iy)^n-\sum_n\frac{\alpha_n}
{2n} (x-iy)^n\right]. \label{6b}
\end{eqnarray}
\end{mathletters}
which is the most general analytic texture with $\hat{c}({\bf}
r=0)=\hat{x}$ and $c_y(x,y)=-c_y(x,-y)$.

The most general form for the surface energy is
\begin{equation}
\sigma(\theta-\Theta)=\sum_{n=0}^{\infty}
a_n\cos\left(n\left(\theta-\Theta\right)\right). \label{7}
\end{equation}
Physically, the coefficient $a_0$ can be identified as the isotropic
line tension. The coefficient $a_1$ measures the lowest order surface
anisotropy. If $a_1 <0$, then the associated surface energy favors a
$\hat{c}$-vector along the outward boundary normal, while for $a_1>0$
it favors a $\hat{c}$ lying along the inward normal. For Langmuir
monolayers, $a_1$ is in general nonzero, but for $D=2$ nematic liquid
crystals $a_1=0$ by symmetry. In that case we must go to the next term
($n=2$). If $a_2<0$ then the surface energy favors a normal
orientation for $\hat{c}$ at the boundary without distinguishong
whether it is outward or inward, while $a_2>0$ favors a parallel
orientation for $\hat{c}$.

Under the assumption of a circular domain, the boundary condition, Eq.
(\ref{3}), becomes
\begin{eqnarray}
\lefteqn{{\kappa}\left[e^{i\theta}f'(e^{i\theta})
-e^{-i\theta}f'(e^{-i\theta}) \right]} \nonumber \\
& & +\frac{R_0}{2}\sum_{n=0}^{\infty} n a_n\left[e^{in\theta}e^{n \left(
-f(e^{i\theta}) + f(e^{-i\theta})\right)} -e^{-in\theta}e^{n \left(
f(e^{i\theta}) - f(e^{ -i \theta})\right)}\right]=0. \label{8}
\end{eqnarray}
Eq. (\ref{8}) can be solved by iteration when $a_n \ll
\kappa/R_0$. For instance, if only $a_0$ and $a_1$ are non-zero, then
we find a Taylor series for $f(z)$:
\begin{equation}
f(z)=-\frac{a_1R_0}{2\kappa}z -
\frac{1}{2}\left(\frac{a_1R_0}{2\kappa}z\right) ^2 +O(a_1^3).
\label{9}
\end{equation}
The choice of the function $f(z)$ is thus imposed by the solution of
Eq. (\ref{3}). Comparing Eqs. (\ref{9}) and (\ref{6}) one can verify
$\alpha_1=-a_1R_0/2 \kappa$ and $\alpha_2=-\left(a_1R_0/2
\kappa\right)^2$. The resulting texture is shown in Fig. 1b. It has a
mathematical singularity {\em outside} the sample, which is called a
``virtual'' boojum. Textures of this type are, indeed, frequently
encountered in both Langmuir monolayers and on the surface of smectic
$C^*$ liquid crystals. The distance of this virtual singularity in the
textural structure from the center of the domain, $R_B$, is obtained
by extrapolating lines of constant $\Theta$ to their point of
intersection. Inserting Eq. (\ref{9}) into Eq.(\ref{4}) and performing
the required extrapolation, one obtains
\begin{equation}
R_B=2 \kappa/a_1R_0  . \label{10}
\end{equation}
Note that as $a_1 \rightarrow 0$, the boojum recedes to infinity and
the texture becomes uniform

For the case $n=2$, the $2D$ nematic drop, we must set $a_1=0$. If
only $a_2$ is non-zero, then interative solution of Eq. (\ref{8})
yields
\begin{displaymath} f(z) \cong -\frac{1}{2}\frac{a_2 z^2}{R_0
\kappa},
\end{displaymath}
so $\alpha_1=0$ and $\alpha_2=-a_2/R_0 \kappa$. The resulting texture
now has {\em two} virtual singularities, but this time they are
located along the $y$ axis.

The above result for $\alpha_1$ and $\alpha_2$, for the case $a_1 \neq
0$, in the perturbation region $a_1 \ll \kappa/R_0$ suggests a Taylor
expansion of $f(z)$ with the coefficients $\alpha_n=\frac{1}{n}\left(
a_1R_0/2 \kappa \right)^n$. This is indeed an exact solution of
Eq.(\ref{8}). In fact, whenever the boundary energy has the special
form $\sigma(\theta - \Theta ) = a_n \cos n (\theta - \Theta)$, i.e.
whenever only one $a_n$  is nonzero (besides $a_0$), an {\em exact}
solution of Eq. (\ref{8}) can be found for $f(z)$:
\begin{equation}
f(z)=\frac{1}{n}\log\left(1-\alpha_nz^n\right), \label{11}
\end{equation}
with
\begin{equation}
\alpha_nR_0^n=\frac{na_nR_0/\kappa}{1+\sqrt{1+\left(na_nR_0/\kappa\right)^2}}.
\label{12}
\end{equation}
A demonstration that Eqs. (\ref{11}) and (\ref{12}) indeed yield a
solution to the boundary condition Eq. (\ref{8}) is contained in
Appendix A.

If $n=1$, then the exact solution corresponds to a virtual boojum,
lying a distance $R_B$ away from the center of the domain, where
\begin{equation}
R_B=R_0\frac{1+\sqrt{1+\left(a_1R_0/\kappa\right)^2}}{a_1R_0/\kappa}.
\label{13}
\end{equation}
As the ratio $a_1R_0/\kappa$ goes to zero, the boojum again retreats
to infinity. As $a_1R_0/\kappa$ goes to infinity, corresponding to a
very strong anisotropic surface energy, or a very large domain,
\mbox{$R_B\rightarrow R_0+\kappa/a_1$}. The boojum thus approaches the
sample in this regime, but the spacing remains finite in the limit
$R_0 \rightarrow \infty$.

When $n$ is greater than one, the exact solution for the texture is
equivalent to the texture produced by $n$ singularities lying outside
of the domain (see Figs. 2b-d). The virtual singularities are no
longer boojums. The singularities for $n>2$ are ``fractionally
charged'' in the sense that we do not recover the starting orientation
of the director field if we perform a circuit around the singularity.
For $n>2$ the resulting singularity is equivalent to a nematic
disclination. There are singular lines attached to the singularities
where $\Theta$ jumps by $2 \pi (n-2)/n$, but these lines do not
intersect the sample.

Textures corresponding to $a_1,\ldots,a_4\neq 0$ are displayed in Fig.
2. For clarity of presentation, the singularities are placed on the
boundary of the domains.

\section{Wulff construction in two dimensions} \label{V}

We now turn to the second part of the problem: minimizing the free
energy with respect to sample shape for a rigid texture. We will first
review the analysis of BCF \cite{Burtonetal} of the $D=2$ Wulff
construction, and then we will use their method to see under what
condtitions we ought to expect to encounter sample shapes with cusps.
The texture is assumed to be uniform in the analysis immediately
following.

We start by introducing a parameterization of a two dimensional curve.
This parameterization expresses the co-ordinates of a point on the
curve in terms of the angle between the tangent to the curve and the
x-axis. If this angle is $\theta$ and the distance between the tangent
line and the origin is $R(\theta)$, then
\begin{mathletters}
\label{w1}
\begin{eqnarray}
x(\theta)&=&R(\theta)\cos(\theta)-\frac{dR(\theta)}{d\theta}\sin(\theta)
\label{w1a} \\
y(\theta)&=&R(\theta)\sin(\theta)+\frac{dR(\theta)}{d\theta}\cos(\theta)
. \label{w1b}
\end{eqnarray}
\end{mathletters}
The quantities $\theta$ and $R(\theta)$ are displayed in Fig. 3. It is
relatively straightforward to express $\theta$ and $R(\theta)$ in
terms of $x$ and $y$. One has
\begin{eqnarray}
\cot(\theta)&=&-\frac{dy}{dx} \nonumber \\
R(\theta)&=&
\left|\frac{x\frac{dy}{dx}-y}{\sqrt{1+\left(dy/dx\right)^2}} \right|.
\label{w2}
\end{eqnarray}
The following relations follow immediately
from Eqs. (\ref{w1}).
\begin{mathletters} \label{w3}
\begin{eqnarray}
\frac{dx}{d\theta}&=&-\sin(\theta)\left[R (\theta)+\frac{d^2R(\theta)}
{d\theta^2}\right] , \label{w3a} \\
\frac{dy}{d\theta}&=&\cos(\theta)\left[R (\theta)+\frac{d^2R(\theta)}
{d\theta^2}\right] ,  \label{w3b}
\end{eqnarray}
\end{mathletters}
and
\begin{equation}
\frac{dS}{d\theta}=\sqrt{\left(\frac{dx}{d\theta}\right)^2 +
\left(\frac
{dy}{d\theta}\right)^2}=\left|R(\theta)+\frac{d^2R(\theta)}
{d\theta^2}\right|. \label{w4}
\end{equation}
Furthermore, the area
inside a closed curve is simply expressed as an integral over
$\theta$:
\begin{equation}
A=\frac{1}{2}\oint\left[xdy-ydx\right]=\frac{1}{2}\oint R(\theta)
\left[R(\theta)+\frac{d^2R (\theta)}{d\theta^2}\right]. \label{w5}
\end{equation}
The integral in Eq. (\ref{w5}) is taken counterclockwise around the
curve.

Suppose, now, that one wishes to minimize the boundary energy of a two
dimensional domain having an anisotropic surface energy,
$\sigma(\theta)$. Here, the angle $\theta$ is both the angle that the
boundary's unit normal makes with respect to the $x$-axis and the
angle that parameterizes the bounding curve in the parameterization of
Eq. (\ref{w1}). This minimization is to be achieved subject to the
constraint that the total enclosed area is a constant. Using Lagrange
multipliers, we arrive at the following extremum equation:
\begin{eqnarray}
0&=& \frac{\delta}{\delta R(\theta)}
\oint\left[\sigma(\theta')
\left[R(\theta')+\frac{d^2R(\theta')}{d\theta'^2}\right] -\frac{
\lambda} {2} R(\theta')
\left[R(\theta')+\frac{d^2R(\theta')}{d\theta'^2}\right]
\right]d\theta' \nonumber \\
&=&\sigma(\theta)+\frac{d^2\sigma(\theta)}{d\theta^2} -
\lambda\left[R(\theta)+\frac{d^2R(\theta)}{d\theta^2}\right],
\label{w6}
\end{eqnarray}
or,
\begin{equation}
R(\theta)+\frac{d^2R(\theta)}{d\theta^2}=\frac{1}{\lambda}
\left[\sigma(\theta)+\frac{d^2\sigma(\theta)}{d\theta^2}\right].
\label{w7}
\end{equation}
The solution of the above equation is
\begin{equation}
R(\theta)=\frac{1}{\lambda}\sigma(\theta)+C_1\cos(\theta)+C_2\sin(\theta).
\label{w8}
\end{equation}
According to Eq. (\ref{w8}), apart from the $C_1$ and $C_2$ terms, the
minimum energy shape has a bounding curve such that $R(\theta)$ is
proportional to the anisotropic surface tension, $\sigma(\theta)$.
Since $R(\theta)$ is the distance from the origin to the tangent of
the bounding curve, we have recovered the Wulff construction
precisely. It can be shown, by direct substitution into Eqs.
(\ref{w1}), that the only effect of the additive sine and cosine terms
is to translate the domain without changing its shape.

As a first example of the use of Eqs. (\ref{w1}) and (\ref{w8}),
consider again the case in which only $a_0$ and $a_1$ are finite in
the fourier expansion, Eq. (\ref{7}), of $\sigma(\theta)$. Setting
$\Theta =0$ in Eq. (\ref{7}) then gives
\begin{equation}
\sigma(\theta)=a_0+a_1\cos(\theta). \label{w9}
\end{equation}
Inserting Eq. (\ref{w9}) into the right hand side of Eq. (\ref{w7}) we
obtain
\begin{eqnarray}
R(\theta)+\frac{d^2R(\theta)}{d\theta^2}&=&
\frac{1}{\lambda}\left[a_0+a_1\cos \theta + \frac{d^2}{d\theta^2}
(a_0+a_1\cos\theta)\right]\nonumber \\
&=&\frac{1}{\lambda}a_0. \label{w10}
\end{eqnarray}
The solution of Eq. (\ref{w10}) is, then, identical to that for the
case $a_1=0$, which is a circle. According to Eq. (\ref{w10}) a simple
$\cos \theta$ contribution to the anisotropic surface energy leads to
{\em no distortion of the shape of the domain}. In fact, we show in
Appendix B that in general a contribution to $\sigma(\theta)$ that is
proportional to $\cos \theta$ cannot affect the sample shape for rigid
textures and, in particular, cannot produce any cusps.

Next, consider a surface energy of the form:
\begin{equation}
\sigma(\theta)=\sigma_0e^{\beta\cos(\theta)}. \label{w11}
\end{equation}
Using Eq. (\ref{w11}) and inserting the result in Eq.
(\ref{w8}) with $C_1=C_2=0$ gives
\begin{mathletters} \label{w12}
\begin{eqnarray}
x(\theta)&=&\left[\cos(\theta)+\beta\sin(\theta)^2\right]e^{\beta\cos(\theta)},
\label{w12a} \\
y(\theta)&=&\left[1-\beta\cos(\theta)\right]\sin(\theta)e^{\beta\cos(\theta)}.
\label{w12b}
\end{eqnarray}
\end{mathletters}
Fig. 4 displays the boundaries for $\beta=0.5,1$ and $1.5$. For
$\beta=1.5$, the bounding curve has a ``swallowtail'' feature.
According to the rule that one keeps only the inner envelope, the
proper prescription is to amputate this feature and retain only the
left-hand portion of the curve, which now has a cusp. The development
of swallowtail singularities in a family of one-parameter curves is a
familiar feature of the theory of catastrophes. It is, for instance,
encountered in a set of curves evolving according to the Huyghens
construction of physical optics, where swallowtail singularities are
associated with the development of caustics. In our case, the curve
parameter is $\beta$, and this parameter must have a finite value for
swallowtail singularities to develop. In Appendix C we prove that the
swallowtail feature appears first at $\beta=1$, so the sample boudary
is smooth for $\beta < 1$.

The key difference between the two anisotropic boundary energies, Eq.
(\ref{w10}) and Eq. (\ref{w11}), is that the latter boundary energy
contains higher order harmonics. Indeed, if we include the higher
harmonics of Eq. (\ref{7}):
\begin{equation}
\sigma(\theta)=\sum_{n=0}^{\infty}a_n\cos(n\theta). \label{w13}
\end{equation}
and insert Eq. (\ref{w13}) into Eq. (\ref{w7})
\begin{equation}
R(\theta)+\frac{d^2R(\theta)}{d\theta^2}=\sum_{n=0}^{\infty}a_n\left(
1 - n^2\right) \cos(n\theta). \label{w14}
\end{equation}
then, as soon as Fourier components with $n>2$ are included in Eq.
(\ref{w14}) we find a distorted boundary, but as long as the $a_n$
coefficients with $n>2$ remain small compared to $a_0$, we find {\em
no} cusp singularity. The swallowtail features cannot in general be
obtained perturbatively. In other words, the anisotropic contribution
to $\sigma(\theta)$ must be comparable to the isotropic line energy
$a_0$ before any singularities in the bounding shape can develop. This
would suggest that the experimentally observed cusp feature can only
be explained if one assumes a highly anisotropic line tension. In the
Section \ref{Robijn} we will use perturbation theory to show why this
conclusion is quite incorrect.

\section{Connection with phase transitions} \label{phase}

The example discussed in the previous Section also serves to
illustrate the mathematical connection between the onset of a cusp and
a thermodynamic phase transition. This connection is more readily
developed if one recasts the problem of calculating the minimum energy
bounding surface into more conventional notation. In this notation the
procedure giving rise to a cusp produces expressions identical to
those encountered in the standard $\phi^4$ model of a
symmetry-breaking phase transition.

We begin by writing the expression for the domain's bounding curve in
the form $x=f(y)$. This single-valued function describes the domain in
the immediate vicinity of $y=0$, where the cusp will occur in the case
at hand. The orientation-dependent surface tension has the form
\begin{eqnarray}
\sigma(\theta)&=&e^{\beta \cos(\theta)} \nonumber \\
&=&e^{\beta/\sqrt{1+\left(dx/dy\right)^2}} \nonumber \\
&=& e^{\beta /\sqrt{1+f'(y)^2}} \label{gl1}
\end{eqnarray}
The surface tension of the portion of the boundary curve lying in the
top half of the $x-y$ plane is
\begin{equation}
\int\sigma(\theta)\,dS = \int_0
e^{\beta/\sqrt{1+f'(y)^2}}\sqrt{1+f'(y)^2}\,dy \label{gl2}
\end{equation}
Taking the functional derivative of the above
expression with respect to $f(y)$, we obtain the following extremum
equation, wich applies at the end-point $y=0$.
\begin{mathletters}
\begin{eqnarray} 0&=& \frac{d}{df'(0)}\left[e^{\beta
\sqrt{1+f'(0)^2}}\sqrt{1+f'(0)^2}\right] \label{gl3a}  \\ &=&
\frac{f'(0)}{1+f'(0)^2}\left[\sqrt{1+f'(0)^2}-\beta \right]
\label{gl3b}
\end{eqnarray}
\end{mathletters}
When $\beta<1$ the only solution to the above equation is $f'(0) = 0$,
while when $\beta>1$ there are the additional solutions corresponding
to $\sqrt{1+f'(0)^2} = \beta$, or, returning to angular variables,
$\beta=1/\cos(\theta_{{\rm cusp}})$.

To gain further insight into the mathematical nature of the onset of
the cusp, we expand the term in brackets on the right hand side of
Eq.(\ref{gl3a}). If $f'(0)$ is small and $\beta \approx 1$, then
\begin{eqnarray} F\left(f'(0)\right) &\equiv&
\sqrt{1+f'(y)^2}e^{\beta/\sqrt{1+f'(0)^2}} \nonumber \\
&=&e^{\beta}\left[ 1+(1-\beta)f'(0)^2 +\frac{1}{8} f'(0)^4 +
O\left(f'(0)^6,(\beta-1)f'(0)^4\right) \right]. \label{gl4}
\end{eqnarray}
The quantity $F\left(f'(0)\right)$ is the surface energy per unit
length along the $y$-axis, at the point the cusp develops.  The right
hand side of Eq. (\ref{gl4}) has precisely the form of a
Ginzburg-Landau-like mean field theory. The combination $1-\beta$
plays the role of the reduced temperature and $f'(0)$ is the order
parameter. The expression possesses a local minimum at $f'(0)=0$ when
$\beta<1$, which becomes a local maximum as $\beta$ passes through
one. The onset of the cusp corresonds to the system's choosing on the
the non-zero values of $f'(0)$ associated with the local minima
at
\begin{equation}
f'(0) = \pm 2 \sqrt{\beta-1} \label{gl5}
\end{equation}
Only one of the solutions of Eq.(\ref{gl5}) is physically relevant.
According to the Wulff construction, the sign of $f'(0)$ must be the
one associated with an outward-pointing cusp. This implies a small
symmetry-breaking term in the theory, at this point unidentified in
our analysis.

In general, if the surface tension $\sigma(\theta)$, has a maximum at
$\theta=0$, then Eq. (\ref{gl4}) can be written as
\begin{equation}
F\left(f'(0)\right) = C_1\left[\sigma(0) +
\sigma''(0)\right]\left(f'(0)\right)^2 + C_2 \left(f'(0)\right)^4.
\label{gl6}
\end{equation}
Cusps appear when $\sigma(0) + \sigma''(0)$ changes sign. This is
consistent with the results of Appendix C, where it is demonstrated
that cusps appear when $R(0) + R''(0)$ changes sign.

\section{Soft limit: effective surface tension} \label{Robijn}

Before developing the general formalism, we will first consider the
limiting case of very small $\kappa$, for which the problem simplifies
significantly. In this limit we can define an effective surface
tension which can be used in the two-dimensional Wulff construction,
as applied to rigid materials. In other words, we include the
deformability of the material by a re-definition of the surface
energy.

In the soft limit the texture of the sample deforms itself to respond
to the surface energy anisotropy. We start by setting $\kappa=0$. The
texture can then adjust itself freely to minimize the anisotropic line
tension. Let $\sigma_0$ be this minimum value. Since the line tension
is a constant, and since the textural energy is zero, the sample shape
must be circular. The line tension is minimized when the director
$\hat{c}$ lies along the outward-directed normal to the circle, so
$\hat{c}$ is in the radial direction along the circle perimeter.  The
associated texture must be a solution of Eq. (\ref{2}) obeying this
boundary condition. In general, the solution will have one or more
singularities. We will focus on the solution of Laplace's equation
with a singularity on the circle perimeter: the boojum texture:
\begin{equation}
\Theta_B(x,y)=2 \arctan \left(\frac{y}{x+R}\right).
\label{r1} \end{equation}

Now, let $\kappa$ be small but finite. Two things must happen: (i) the
$\hat{c}$-director exerts a torque on the boundary which is then
deformed away from a perfectly circular shape, and (ii)  the
$\hat{c}$-director is no longer perfectly along the boundary normal.
We will first keep the texture fixed at the at Eq. (\ref{r1}),
allowing the sample shape to relax, and then we will allow the texture
to relax and reconsider the sample shape.

First, use Gauss's law to rewrite the total energy as a line integral
over the boundary assuming $\Theta$ to be a solution of Eq. (\ref{2}):
\begin{equation}
F=\oint
ds\left\{\frac{\kappa}{2}\Theta(s)\frac{\partial \Theta (s)}{\partial
n} + \sigma\left(\theta-\Theta\right)\right\}. \label{r2}
\end{equation}
The quantity $\sigma_{{\rm eff}}$ given by
\begin{equation}
\sigma_{{\rm
eff}}=\frac{\kappa}{2}\Theta\frac{\partial \Theta}{\partial n} +
\sigma\left( \theta - \Theta \right)   \label{r3}
\end{equation}
then appears as an effective line tension. The function $\Theta(s)$
is, however, a functional of the boundary shape so the right hand side
of Eq. (\ref{r3}) is really a non-local expression and cannot be
simply interpreted as a line tension. If, nevertheless, we use  use
Eq. (\ref{r1}) in Eq. (\ref{r3}) we find: \begin{equation}
\sigma_{{\rm eff}}=\frac{\kappa}{2 R} \theta \frac{\sin \theta}{1-\cos
\theta} + \sigma_0. \label{r4}
\end{equation}
The resulting effective surface tension is an analytic, but aperiodic
function of $\theta$. The line tension anisotropy only depends on the
dimensionless parameter $\Gamma=\kappa / \sigma_0 R$, with $R$ the
sample radius. It can be inserted into the Wulff construction and
leads to a cusp at $\theta=0$, with a cusp angle proportional to
$\Gamma$.

We now redo this calculation while allowing for textural relaxation.
First, assume a circular boundary. Allow $\Theta$ to deviate from the
boojum texture, and expand the line energy around its minimum
$\sigma_0$:
\begin{eqnarray}
\Theta(x,y) &=& \Theta_B(x,y) + w(x,y)
\label{r5}\\
\sigma \left( \theta-\Theta \right) &\approx& \sigma_0 +
\frac{1}{2} \sigma_0''\left( \theta -\Theta\right)^2. \label{r6}
\end{eqnarray}
To find $w$ we must solve $\nabla^2w=0$ subject to the
boundary conditions Eq. (\ref{3}). In terms of $w$:
\begin{equation}
\kappa\left. \frac{\partial w}{\partial r}\right|_{r=R} +\sigma ''(0)
w(s) = -\kappa \left. \frac{\partial \Theta_B}{\partial
r}\right|_{r=R} . \label{r7}
\end{equation}
The solution of this mixed boundary condition problem for $w$ can be
written as a Fourier expansion in $\theta$:
\begin{equation}
w(r, \theta) = \sum_{m=1}^{\infty}a_m r^m \sin m \theta \label{r8}
\end{equation}
with coefficients:
\begin{equation}
a_m = -\frac{\kappa}{\pi \sigma
''(0) R_m\left(1+\frac{\kappa m}{\sigma''(0)}\right) }
\int_{-\pi}^{\pi} d\theta '\left.\frac{\partial \Theta_B(\theta
')}{\partial r} \right|_{r=R}. \label{r9}
\end{equation}
Inserting Eqs. (\ref{r8}) and (\ref{r9}) into Eq. (\ref{r2}) yields:
\begin{equation}
F=R\int_{-\pi}^{\pi} d\theta \left\{\sigma_0 -
\frac{1}{2}\sigma_0''\theta w(r=R,\theta) + O(w^4)\right\} ,
\label{r10}
\end{equation}
where we have used Eq. (\ref{r7}). We neglect the fourth and higher
order terms in $w$. Inserting the solution Eq. (\ref{r8}) into the
line integral expression for $F$ gives, as expected, a non-local
expression:
\begin{equation}
F=R\int_{-\pi}^{\pi}d\theta \left \{ \sigma_0 + \theta
\int_{-\pi}^{\pi} d\theta' K(\theta, \theta')\left. \frac{\partial
\Theta_B(\theta')}{\partial r} \right|_{r=R} \right \},\label{r11}
\end{equation}
with a kernel $K(\theta, \theta')$:
\begin{equation}
K(\theta,\theta')=\frac{\kappa}{2 \pi}\sum_{m=1}^{\infty}\frac{\sin m
\theta \sin m \theta'}{1+\frac{m \kappa}{\sigma_0 ''R}}. \label{r12}
\end{equation}
In the large $R$ limit, this kernel reduces to two delta functions, at
$\theta=\pm \theta'$, while for finite $R$, these two delta functions
broaden by an amount $\delta \theta$ of order $\kappa/\sigma''R$. To
second order in $\kappa/R$ we can neglect the spreading of the delta
functions. In the large $R$ limit, the non-local line energy thus
becomes a {\em local} line tension. Using Eq. (\ref{r12}) in
(\ref{r11}), one obtains:
\begin{eqnarray}
F&=&R\int_{-\pi}^{\pi} d\theta \left \{ \sigma_0 + \frac{\kappa}{2}
\theta \left. \frac{\partial \Theta_B(\theta)}{\partial
r}\right|_{r=R}\right\}. \label{r13} \\ &&R\rightarrow \infty
\nonumber
\end{eqnarray}
Surprisingly, we recover the ``naive'' line tension of Eq. (\ref{r4}).
If we now allow the shape to relax, we find the shape discussed
earlier. This shape relaxation must then be used in a recalculation of
$w$, but in the large $R$ limit such corrections are higher order in
$1/R$. We conclude that in the limit of small $\kappa$ and large $R$
the sample shape is expected to have a cusp with an excluded angle
proportional to $\kappa/\sigma_0R$.

\section{Generalization of the Wulff construction} \label{VII}

We now relax the condition of small $\kappa$ and develop our method,
which allows for a determination of the sample shape, even when the
$\sigma_{{\rm eff}}$ of Eq. (\ref{r3}) is truly non-local. Our method
will be a generalization of the BCF procedure\cite{Burtonetal} and the
results of Section \ref{III}. We will assume that the texture always
obeys Laplace's equation, Eq. (\ref{2}), with the director angle
$\Theta$ specified by a complex function $f(z)$ through Eq. (\ref{4}).
The function $f(z)$ is, for a given sample shape, determined by Eq.
(\ref{8}). The remaining problem is now to minimize the sample free
energy with respect to sample shape. We will use the BCF
parameterization $R(\theta)$ for the sample shape. To find the optimal
sample shape, we must compute the variational derivative of both
surface and textural energy and equate it to the variational
derivative of the sample area $A$ with respect to $R(\theta)$ times
the Lagrange multiplier $\lambda$---in direct analogy with Eq.
(\ref{w6}). We will, in the following, always assume near-circular
shapes.

We start with the variational derivative of the textural energy.
Imagine a two dimensional domain containing an order parameter
described by the director angle $\Theta(x,y)$. If $\Theta(x,y)$ is
given by Eq. (\ref{4}) then
\begin{equation}
\left|\vec{\nabla}\Theta \right|^2 = 4 f'(x+iy)f'(x-iy) ,
\label{gw1}
\end{equation}
and the bulk contribution to the energy in Eq. (\ref{1}) is, then,
\begin{equation}
2 \kappa \int F(x,y(x))dx , \label{gw2}
\end{equation}
where $y(x)$ denotes the boundary line energy, and where
\begin{equation}
F(x,y) = \int^yf'(x+iy')f'(x-iy')\,dy' .
\label{gw3}
\end{equation}
The textural energy, Eq. (\ref{gw2}) is explicitly dependent on the
shape of the boundary, and, again, can be treated as a non-local
contribution to the line tension. Using Eqs. (\ref{w3}) and (\ref{w4})
we rewrite the infinitesimal $dx$ as $-\sin \theta ds$, with
$s(\theta)$ the arclength. The derivative with respect to $R(\theta)$
of the textural energy Eq. (\ref{gw2}) is then
\begin{eqnarray}
\lefteqn{\frac{\delta}{\delta R(\theta)} 2 \kappa
\int F\left( x\left(\theta'\right),y\left( \theta' \right) \right)
\sin \theta'\,ds(\theta')} \nonumber \\  &=&2 \kappa
f'\left(x\left(\theta \right)+iy\left( \theta \right)\right)
f'\left(x\left(\theta  \right)-iy\left( \theta \right)\right).
\label{gw4}
\end{eqnarray}

We now turn to the variational derivative of the surface energy, Eq.
(\ref{7}). Using the Fourier expansion, Eq. (\ref{7}), for
$\sigma(\theta-\Theta)$ we find:
\begin{eqnarray}
\lefteqn{\frac{\delta}{\delta R(\theta)}
\int\sigma\left(\theta-\Theta\right)\,dS =} \nonumber \\ && {\rm Re}
\left\{\sum_n a_n e^{-in\theta} e^{nf(x + i y)-nf(x-iy)}\left[ n(n+1)(
x+iy)f'(x + i y) + n(n-1) (x-i y) f'(x-i y) +1-n^2 \right]\right\}.
\nonumber \\ \label{gw5}
\end{eqnarray}
The algebraic steps in the derivation of Eqs. (\ref{gw4}) and
(\ref{gw5}) are outlined in Appendix D.

The two variational derivatives appear to be forbiddingly intricate.
However, our complex function $f(z)$ is not just any function; it must
obey Eq. (\ref{8}), and we use this to simplify Eqs. (\ref{gw4}) and
(\ref{gw5}). First, we rescale lengths in the circular domain so that
its radius is unity, as described in Section \ref{III}. On the
boundary of the domain, one can then replace $x+iy$, respectively
$\exp(in\theta)$ by $z$, respectively $z^n$ (where $|z|=1$) and the
quantities $x-iy$, respectively $\exp(-in\theta)$ by $1/z$,
respectively $1/z^n$. Equation (\ref{8}) which determines $f(z)$ then
simplifies to:
\begin{equation}
\kappa\left[zf'(z)
-\frac{1}{z}f'(\frac{1}{z}) \right] +\frac{R_0}{2}\sum_n n
a_n\left[z^ne^{n \left( -f(z) + f(1/z)\right)} -z^{-n}e^{n \left( f(z)
- f(1/z)\right)}\right] =0. \label{gw6}
\end{equation}.

Dividing both sides by $z$ and integrating yields
\begin{eqnarray}
\kappa\int^z\left[f'(x)-\frac{1}{x^2}f'\left(\frac{1}{x}\right)
\right] dx&=& -\sum_n\frac{a_n}{2}\left[z^ne^{-nf(z)+nf(1/z)} +
z^{-n}e^{nf(z)-nf(1/z)}\right] \nonumber \\
&&+\kappa\int^z\left[xf'(x)-\frac{1}{x}f'\left(\frac{1}{x}\right)\right]\left[f'(x)
+\frac{1}{x^2}f'\left(\frac{1}{x}\right)\right]\,dx \label{gw7}
\end{eqnarray}

Using Eqs. (\ref{gw6}) and (\ref{gw7}) in Eqs. (\ref{gw4}) and
(\ref{gw5}) we find that the functional derivative with respect to
$R(\theta)$ of the energy of the domain, as given by the sum of the
right hand sides of Eqs. (\ref{gw4}) and (\ref{gw5}), is equal to
${\cal F}(z)$, with
\begin{eqnarray}
{\cal F}(z)&=& \nonumber \\ &&z
\kappa\frac{d}{dz}\left[zf'(z)-\frac{1}{z}f'\left(\frac{1}{z}\right)\right]
+\kappa\left[zf'(z)\right]^2+
\kappa\left[\frac{1}{z}f'\left(\frac{1}{z}\right)\right]^2 \nonumber
\\ &&-\kappa\left[f(z)+f\left(\frac{1}{z}\right)\right] +
\kappa\int^z\left[wf'(w)^2 -
\frac{1}{w^3}f'\left(\frac{1}{w}\right)^2\right]\,dw , \nonumber \\
\label{gw8}
\end{eqnarray}
up to an unimportant constant.

The equation determining the optimal shape is now found by repeating
the steps between Eqs. (\ref{w5}) and (\ref{w6}), with the result:
\begin{equation}
R(\theta)+\frac{d^2R(\theta)}{d\theta^2}=\frac{1}{\lambda}\left[\sigma_0R_0
+ {\cal F}(\theta)\right], \label{gw9}
\end{equation}
where $\sigma_0$ is the isotropic contribution to the surface tension,
and where ${\cal F}(\theta)$ is given by Eq. (\ref{gw8}) with $z$
replaced by $e^{i\theta}$.

Equations (\ref{gw8}) and (\ref{gw9}), which hold as long as the
domain boundary is not too far deformed from a circular shape, form
the basis of our method. The sample shape is determined in three
steps: (i) solve Eq. (\ref{gw6}) to obtain $f(z)$ for a given
anisotropic line tension $\sigma(\theta-\Theta)$, (ii) compute ${\cal
F}(\theta)$ from Eq. (\ref{gw8}) to obtain the effective line tension,
and (iii) solve the BCF formula, Eq. (\ref{gw9}) with ${\cal
F}(\theta)$ to find $R(\theta)$. We will now consider some examples of
the application of our method.

\subsection{ $\sigma(\phi)=\sigma_0+\lowercase{a}_1\cos
\phi$}\label{VIII}

Our first example is the case in which $a_n=0$ for $n>2$ in the
Fourier expansion, Eq. (\ref{7}). Recall that for an imposed, rigid,
uniform texture the minimum energy shape was a perfect circle (Eq.
(\ref{w10})), and that for an imposed circular domain shape the
minimum energy texture was the virtual boojum (Eq. (\ref{11}) with
$n=1$). To implement our recipe for finding the shape which minimizes
the total free energy, we first note that the virtual boojum texture:
\begin{equation}
f(z)=\log(1-\alpha_1z) \label{gw10}
\end{equation}
is an exact solution of Eq. (\ref{gw6}), with $\alpha_1$ given by Eq.
(\ref{12}). If we then insert $f(z)$ into ${\cal F}(z)$ (Eq.
(\ref{gw8})) we find ${\cal F}(z) = {\rm constant}$. The resulting
shape equation
\begin{equation}
R(\theta)+\frac{d^2R}{d\theta^2}=\frac{1}{\lambda}\left[\sigma_0R_0+{\rm
const.}\right]   \label{gw11}
\end{equation}
is the equation for a perfect circle. {\em We conclude that the
virtual boojum texture in a circular domain is a free energy minimum,
both with respect to domain shape and domain texture}. We have yet to
find a simple physical argument that explains why the perfect circle
remains a free energy minimum for this anisotropic form of the line
tension.

\subsection{$\sigma(\phi)=\sigma_0+\lowercase{a}_{\lowercase{n}} \cos
\lowercase{n} \phi\,\,\,(\lowercase{n}>1)$}

Let us now try the same procedure for $n>1$. First, assume a perfect
circular sample. The associated texture then follows from
$f(z)=\frac{1}{n} \log(1-\alpha_nz^n)$, which is a solution of Eq.
(\ref{gw6}). If we now compute ${\cal F}(z)$ and insert the result in
the shape equation we find
\begin{eqnarray}
\lefteqn{R(\theta)+\frac{d^2R(\theta)}{d\theta^2}}\nonumber \\ &=& R_0
+ \frac{\kappa}{\sigma_0}
\left[\frac{(n-1)^2}{n}\frac{1}{1-\alpha_ne^{in\theta}}+
\frac{1-n}{(1-\alpha_ne^{in\theta})^2}\right] \nonumber \\ &&+
\frac{\kappa}{\sigma_0}
\left[\frac{(n-1)^2}{n}\frac{1}{1-\alpha_ne^{-in\theta}}+
\frac{1-n}{(1-\alpha_ne^{-in\theta})^2}\right]. \label{gw12}
\end{eqnarray}
The circular shape is no longer a free energy minimum.
In the limit of very small domains, with $\sigma_0R_0/\kappa \ll 1$,
Eq. (\ref{gw11}) reduces to
\begin{equation}
R(\theta)+\frac{d^2R(\theta)}{d\theta^2} \simeq \frac{R_0}{\sigma_0}
\left[ \sigma_0 +(1-n^2) a_n \cos n \theta\right], \label{gw13}
\end{equation}
with solution
\begin{equation}
R(\theta) \cong
R_0\left(1+\frac{a_n}{\sigma_0} \cos n\theta \right). \label{gw14}
\end{equation}

The domain shape has an $n$'th harmonic shape deformation on top of
the circular shape. Note that the domain shape is independent of the
Frank constant $\kappa$. This result is, of course, not quite exact,
since $f(z)$ will, for a non-circular shape, no longer be given by Eq.
(\ref{11}), but the resulting corrections are higher order in
$a_n/\sigma_0$ and $\sigma_0R_0/\kappa$.

\subsection{$\sigma(\phi)=\sigma_0+a_1 \cos \phi + a_2 \cos 2 \phi$}

We have seen under A that for a pure $\cos \phi$ line tension
anisotropy we can find a free energy minimum in the perfectly circular
shape. Since we have an exact solution, we can use perturbation theory
to asses the effect of higher order harmonics. We thus include one
more term---$n=2$, with $a_2 \ll a_1$---and recompute the shape. This
particular form of the surface term has been argued to be relevant for
Langmuir films \cite{Robijn}. From our earlier results one would
naively expect a nearly circular sample shape with a small $n=2$
correction. However, things turn out a little differently.

If we take $a_2\ll a_1$ then it will be possible to expand $f(z)$
about its $a_2=0$ form. Writing
\begin{equation}
f(z)=\log(1-\alpha z)+f_1(z), \label{gw15}
\end{equation}
with $\alpha\equiv \alpha_1 R_0$ given by Eq. (\ref{12}), the boundary
condition Eq. (\ref{gw6}) becomes, to first order in $a_2$ and $f_1$,
\begin{eqnarray}
\lefteqn{\frac{\kappa}{R_0}\left[zf_1'(z)-\frac{1}{z}f_1'\left(\frac{1}{z}\right)
\right] -\frac{a_1}{2}f_1(z)\left[\frac{z-\alpha}{1-\alpha
z}+\frac{1-\alpha z}{z-\alpha}\right]}\nonumber \\
&&+\frac{a_1}{2}f_1\left(\frac{1}{z}\right)\left[\frac{z-\alpha}{1-\alpha
z} + \frac{1-\alpha z}{z-\alpha}\right]
+\frac{a_2}{2}\left[\left[\frac{z-\alpha}{1-\alpha
z}\right]^2-\left[\frac{1-\alpha z}{z-\alpha}\right]^2\right]
\nonumber \\ &&=0. \label{gw16}
\end{eqnarray}
We separate Eq. (\ref{gw16}) into two equations:
\begin{equation}
\frac{a_1}{2}\left[\frac{1}{\alpha}-\alpha\right]zf_1'(z)
-\frac{a_1}{2}f_1(z)\left[\frac{z-\alpha}{1-\alpha z}+\frac{1-\alpha
z}{z-\alpha} \right] +\frac{a_2}{2}\left[\frac{z-\alpha}{1-\alpha
z}\right]^2 = 0, \label{gw17}
\end{equation}
and an identical equation with $z$ replaced by $1/z$. Each of those
equations is a simple linear first order differential equation,
solvable by standard methods. One finds
\begin{equation}
f_1(z)=-\frac{a_2}{a_1}\frac{\alpha}{1-\alpha^2}\frac{z-\alpha}{1-\alpha
z} \int^1 t^{2\alpha^2/(1-\alpha^2)}\frac{zt-\alpha}{1-\alpha z
t}\,dt. \label{gw18}
\end{equation}

Computing ${\cal F}(z)$ with the use of $f(z)=\log(1-\alpha z)+f_1(z)$
one finds the shape equation
\begin{equation}
r(z)= \frac{1}{2}
-\frac{\kappa}{\sigma_0R_0}\left[f_1(z) + 2 \alpha(1-\alpha
z)\int^z\frac{f_1(z)} {(1-\alpha z)^2}dx\right], \label{gw19}
\end{equation}
where $R(z)=r(z)+1/r(z)$. The final reduction of the result for
$R(\theta)$ consists of algebraic manipulations. First, we note that
the quantity $\alpha \equiv \alpha_1 R_0$ as given by Eq. (\ref{12})
approaches $1$ as $R_0 \rightarrow \infty$. This means that, in the
limit that $R_0$ is large we can replace $1-\alpha$ by
$\kappa/a_1R_0$. Furthermore, inspection reveals that in the limit
$\kappa/a_1R_0\ll 1$ and $a_2\ll a_1$ the first order contribution to
$x(\theta), y(\theta)$ is negligible compared to the zeroth order one
except in the immediate vicinity of $\theta=0$. After a series of
changes of variables that take this behavior into account we arrive at
the following expression for $R(\theta)$
\begin{equation}
R(\theta)=1+\frac{2\kappa}{\sigma_0 R_0}\frac{a_2}{a_1}{\rm
Re}\left[{\rm {\bf X}}\left(\frac{\theta R_0
a_1}{\kappa}\right)\right] , \label{gw20}
\end{equation}
where the dimensionless function ${\rm{\bf X}}(\psi)$ is given by
\begin{eqnarray}
\lefteqn{{\rm{\bf X}}(\psi)} \nonumber \\
&=&\frac{1+i\psi}{1-i\psi}\int_0^{\infty}\frac{e^{-y}}{1-i\psi+y}dy
\nonumber \\
&&-4(1-i\psi)\int_0^{\infty}\frac{e^{-y}}{y^3}\left\{\log(1-i\psi+y)
-\log(1-i\psi) -\frac{y}{1-i\psi} +\frac{y^2}{2(1-i\psi)^2}\right\}dy
\nonumber \\
&&+2(1-i\psi)\int_0^{\infty}\frac{e^{-y}}{y^2}\left\{-\log(1-i\psi+y)+\log(1-i\psi)
+ \frac{y}{1-i\psi}\right\}dy. \label{gw21}
\end{eqnarray}

A graph of the real part of the function ${\rm{\bf X}}(\psi)$ is
displayed in Fig. 5. The following properties of the function play an
important role in the behavior of the boundary
\begin{enumerate}
\item{${\rm Re}\left[{\rm{\bf X}}(\psi)\right]$ is an even,
nonsingular function of its argument, and can thus be expanded in an
power series in $\psi$.}
\item{${\rm Re}\left[{\rm{\bf X}}(\psi)\right]$ has a negative second
derivative at $\psi=0$.} \item{${\rm Re}\left[{\rm{\bf
X}}(\psi)\right]$ possesses a minumum at $\psi\approx2.74$.}
\end{enumerate}

\subsubsection{Onset of a cusp}

As discussed in Section \ref{phase} and Appendix C, the onset of a
cusp is signaled by a change in sign of the function
$R(\theta)+d^2R(\theta)/d\theta^2$. From the previous arguments it
follows that this is most likely to happen at $\theta=0$. Making use
of our result, Eq. (\ref{gw20}), for $R(\theta)$ we find
\begin{equation}
\left. R(\theta)+\frac{d^2R(\theta)}{d\theta^2}\right|_{\theta=0}=1+
\frac{2\kappa}{\sigma_0 R_0} \frac{a_2}{a_1}{\rm Re}{\rm {\bf X}}(0) +
\frac{2a_2a_1R_0}{\sigma_0 \kappa}{\rm {\bf X}}''(0). \label{gw22}
\end{equation}

In the limit of vary small $a_2$ the  expression on the right hand
side will be positive unless the domain radius $R_0$ exceeds a
threshold value, the leading contribution to which is given by
\begin{eqnarray}
R_{{\rm threshold}}&=&\frac{\sigma_0 \kappa} {2 a_1 a_2 \left|{\rm
{\bf X}}''(0) \right|} \nonumber \\ &=&0.356 \frac{\sigma_0
\kappa}{a_1 a_2}, \label{gw23}
\end{eqnarray}
where the number in the last equality above follows from a numerical
evaluation of the second derivative. When the sample radius is
slightly in excess of $R_{{\rm threshold}}$ the excluded cusp angle,
$\psi$ (the difference between the cusp angle and $2 \pi$), obeys
\begin{equation}
\psi \propto \sqrt{R-R_{{\rm threshold}}}. \label{gw24}
\end{equation}
As in the case of the example discussed in Section \ref{phase}, the
behavior of the cusp angle at onset obeys a power law consistent with
a $\phi^4$ mean field theory.  A graph of the excluded angle of the
cusp as a function of domain radius can be found in Fig. 6. The domain
radius is expressed in units of $\kappa/a_1$. In that case, the only
other adjustable parameter is the ratio $a_2/\sigma_0$. In the Figure,
that ratio is equal to $0.05$.

\subsubsection{Large $R$ limit}

The behavior of the excluded angle when $R_0$ is very large can also
be determined by inspection of Eq. (\ref{gw20}). As the symmetry of
the domain is the same in this case as in the illustrative case
discussed in Section \ref{V} and in Appendix C, we locate a cusp by
searching for a solution to the equation $y(\theta)=0$, with $\theta$
small but not equal to zero. Substituting the right hand side of Eq.
(\ref{gw20}) into Eq. (\ref{w1b}) we find that, for small $\theta$ and
large $R_0$, with $\theta R_0$ finite,
\begin{equation}
y(\theta) = \frac{2 a_1a_2R_0}{\sigma_0 \kappa} \theta
\left\{\frac{{\rm Re}\left[{\rm {\bf X}}'\left( \theta R_0
a_1/\kappa\right)\right]}{\theta R_0 a_1/\kappa}\right\}. \label{gw25}
\end{equation}

Because of the properties of the function ${\rm {\bf X}}(\psi)$
enumerated above, the ratio in curly brackets in Eq. (\ref{gw25}) is
well-behaved as a function of the variable $\theta R_0 a_1/\kappa$.
Neglected in Eq. (\ref{gw25}) are terms of zeroth order in $R_0$.

According to Eq. (\ref{gw25}), $y(\theta)=0$ when either $\theta=0$ or
 ${\rm Re}\left[{\rm {\bf X}}'\left( \theta R_0
a_1/\kappa\right)\right]=0$. Finite cusp angles correspond to the
second case. Because ${\rm Re}\left[{\rm {\bf X}}(\psi)\right]$ when
$\psi=2.74$, we conclude that $\theta = 2.74\left(
\kappa/R_0a_1\right)$. The interior angle of the cusp is then equal to
$\pi-5.48 \kappa/a_1 R_0$ (see Appendix C) and the excluded angle of
the cusp, defined as $\pi$ minus the interior angle, is equal to $5.48
\kappa/a_1 R_0$. The excluded angle decays as $1/R_0$ and is {\em
independent} of $a_2$ when $R_0$ is sufficiently large.

Fig. 7 shows what the domain boundary looks like when a cusp is
induced by the $a_2 \cos 2 \phi$ term. The energy parameters have been
set so that the inequalities relied upon in this Section are
satisfied. In particular, $a_2/\sigma_0=0.25$ and $R_0a_1/\kappa=5$.
Note that a swallowtail appears, just as in the illustrative example
cited previously.

\section{conclusions and comparison with experiment}

In summary, we have demonstrated that a two-dimensional domain
containing a deformable medium describable by an $X-Y$ model has a
purely circular boundary, as long as the surface anisotropy term is of
the form $a_1 \cos \phi$, where $\phi$ is the angle between the $X-Y$
vector and the unit normal to the bounding curve. This remarkable
result no longer holds if one introduces the small anisotropic surface
energy $a_2 \cos 2 \phi$. In certain regimes the surface deformation
is singular; a cusp appears when the sample size exceeds a threshold
determined by the coefficients of the anisotropic and the bulk Frank
constant, $\kappa$. The cusp angle ultimately decays with increasing
sample size as one over the domain radius. The threshold domain radius
is proportional to $\sigma_0 \kappa /a_1 a_2$, where $\sigma_0$ is the
isotropic bounary energy. Note that the more deformable the material
in the domain is (i.e. the smaller the value of $\kappa$) the more
likely it is to develop a cusp---which contradicts  naive intuition
based on the appearance of facets in crystalline materials.

How does our theory compare with the experimentally measured cusp
angles? In Fig. 8 we show a set of cusp angles measured as a function
of sample size by Schwartz, et al \cite{KnoblerandSchwartz} for
pentadecanoic acid. The data has been fitted to the curve displayed in
Figure 6. The ratio $a_2/\sigma$ has been set equal to $0.05$. No
attempt was made to adjust that ratio so as to optimize the fit, but
the quatities $a_1$ and $\kappa$ were effectively varied by adjusting
the vertical and horizontal scales. Note that the measured cusp angles
show no evidence of a sharp onset at a threshold radius.

An even more serious discrepancy between the data and our results lies
in the fact that the best fit to the asymptotic power law decay of the
measured cusp angle with domain size is $\theta_{\rm cusp} \propto
R_0^{-0.3}$. Our mean field theory gives a decay as $1/R_0$. One
possible origin of the discrepancy is that we did not identify the
correct operator as the one that ``breaks'' the degeneracy of the pure
$a_1 \cos \psi$ boundary energy. Langmuir monolayers have Frank
constants $\kappa_1$ and $\kappa_3$ for splay and bend that are not
identical \cite{Robijn}. We have assumed in Eq. (\ref{1}) that
$\kappa_1=\kappa_3$. The bulk equation satisfied by the director angle
$\Theta$ is now nonlinear. Is the solution to this equation related in
any way to the virtual boojum solution that plays such an important
role in the determination of the domain's shape? Preliminary work
indicates that cusps also appear if we let $\kappa_1 \neq \kappa_3$.
Another possibility is that thermal fluctuations lead to a
renormalization of the effective anisotropic boundary energies
\cite{Saleur}. This possibility is under current study.

The cusp singularities discussed in this paper should be generic
features of boundary lines in Langmuir layers and two-dimensional
liquid crystalline materials in general. Indeed, defect lines in
hexatic Langmuir monolayers frequently exhibit a scalloped appearance
\cite{Robijn}, which may have the same origin as the cusps discussed
here. The extension of the work reported here to more general sample
shapes is not straightforward. with regard to the family of exact
solutions for the texture we have found, an obvious question is
whether some variation on a conformal transformation allows one to
generate from it the texture appropriate to a non-circular domain. It
is readily verified that a simple conformal transformation will not
simultaneously deform the domain's bounding curve and appropriately
alter the texture. This is because of the nonlinear nature of the
boundary conditions. However, given that we have obtained a set of
exact solutions for the texture in the presence of realistic boundary
conditions, an extension of the applicability of those solutions would
be highly desirable.

\section{acknowledgements}

We are grateful to Professor Charles Knobler and Daniel Schwartz for
extremely useful discussions, including the key suggestion by
Professor Schwartz that led to this work. We would also like to thank
them for generously sharing their data with us at all stages. We would
also like to thank J. Meunier  and P. Nelson for helpful discussions.
R. B. was supported by DMR grant DMR 9407747.

\appendix \section{The virtual boojum solution}

In this Appendix we show that Eqs. (\ref{11}) and (\ref{12}) represent
exact solutions to the boundary value problem represented by Eq.
(\ref{8}), with one nonzero $a_n$. The demonstration is by direct
substitution. We have
\begin{eqnarray}
\lefteqn{\frac{1}{2}a_n\left[e^{in\theta}e^{n \left( -f(e^{i\theta}) +
f(e^{-i\theta})\right)} -e^{-in\theta}e^{n \left( f(e^{i\theta}) -
f(e^{ -i \theta})\right)}\right]}\nonumber \\ & & =
\frac{1}{2}a_n\left[\frac{e^{in\theta}-\alpha_n }{1-\alpha_n
e^{in\theta}} - \frac{e^{-in\theta}-\alpha_n }{1-\alpha_n
e^{-in\theta}}\right] \nonumber \\ & & =\frac{1}{2}
a_n\left[\frac{1/\alpha_n-\alpha_n }{1-\alpha_n e^{in\theta}} -
\frac{1/\alpha_n-\alpha_n }{1-\alpha_n e^{-in\theta}}\right]
\label{a1}
\end{eqnarray}
and
\begin{eqnarray}
\frac{\kappa}{R_0}\left[e^{i\theta}f'(e^{i\theta})
-e^{-i\theta}f'(e^{-i\theta}) \right] &=&
\frac{\kappa}{R_0}\left[-\frac{\alpha_n
e^{in\theta}}{1-\alpha_ne^{in\theta}} + \frac{\alpha_n
e^{-in\theta}}{1-\alpha_ne^{-in\theta}}\right] \nonumber \\
&=&\frac{\kappa}{R_0}\left[-\frac{1}{1-\alpha_ne^{in\theta}} +
\frac{1}{1-\alpha_ne^{-in\theta}}\right] .\nonumber \\ \label{a2}
\end{eqnarray}
According to Eq.(\ref{8}), the sum of the left hand
sides of Eqs. (\ref{a1}) and (\ref{a2}) equals zero. The two right
hand sides sum to zero if
\begin{equation}
\frac{1}{\alpha_n}-\alpha_n-\frac{2 \kappa}{a_nR_0}=0. \label{a3}
\end{equation}
The solution to this equation for the parameter $\alpha_n$ that has
the proper limiting behavior is displayed in Eq. (\ref{12}).

\section{Stability of the boundary against $\sigma(\theta)\propto
\cos(\theta)$}

Consider an anisotropic surface tension having the form
$\sigma(\theta)=A\cos(\theta)$. The total boundary energy of a closed
boundary of arbitrary shape is
\begin{equation}
\oint \hat{c}\cdot \hat{n} ds. \label{b1}
\end{equation}
The quantity $\hat{c}$ in the above expression is a constant vector,
$\hat{n}$ is the unit normal to the boundary and $ds$ is the
infinitesimal element of length along the boundary. In two dimensions
we can write
\begin{equation}
\hat{c}\cdot \hat{n} ds= \hat{z} \cdot \left(\hat{c} \times
d\vec{s}\right), \label{b2}
\end{equation}
where $\hat{z}$ is the unit vector out of the plane and $d\vec{s}$ is
the directed infinitesimal length element along the boundary. This
means that the boundary energy is
\begin{equation}
\hat{z} \cdot\left[\left(\oint d\vec{s}\right)\times
\hat{c}\right]. \label{b3}
\end{equation}

The closed integral on the right hand side of Eq. (\ref{b3}) is always
equal to zero. This means that a $\sigma(\theta)\propto \cos(\theta)$
has {\em absolutely no effect} on the shape of a two dimensional
domain.

\section{On the appearance of cusps}

To demonstrate exactly how the implementation of the Wulff
construction in two dimensions leads to cusps on boundaries we
consider the equation for the bounding curve generated by Eqs.
(\ref{w12})---that is, the boundary when the anisotropic surface
tension is given by Eq. (\ref{w11}). Because of the symmetry of the
domain, as illustrated in Figs 4, a cusp lying on the rightmost edge
of the domain represents a solution to the equation $y(\theta)=0$,
subject to the additional condition $\theta \neq 0$. According to Eq.
(\ref{w12b}), the above requirements are met when
\begin{equation}
\beta \cos\theta=1. \label{c1}
\end{equation}
This equality can only be satisfied if $\beta >1$. By simple geometry,
we find that the interior angle of the cusp is equal to $\pi-2
\theta_c$, where $\theta_c = \arccos(1/\beta)$ is the angle satisfying
Eq. (\ref{c1}).

Fig. 4c depicts the boundary when $\beta=1.5$, while Fig. 4b show how
the domain looks when $\beta$ is equal to its critical value of $1$.

Of interest is the ``swallowtail'' appendage attached to the cusp in
the former case. The construction of the inner envelope corresponding
to the equilibrium boundary shape is completed by the amputation of
this appendage. Although the behavior of $x(\theta)$ and $y(\theta)$
in the tail is, thus, nominally irrelevant to the shape of the domain,
a brief discussion of the properties of $x$ and $y$ in the tail region
yields useful information regarding the mathematical signals that
accompany the onset of a cusp. Consider, for instance, the cusps in
the tail, at either end of the ``trailing'' edge. Using Eqs.
(\ref{w3}), we find $dy/dx = -\cot (\theta)$. Thus, there will be no
discontinuity in the slope of the curve given by Eqs. (\ref{w1}) as
long as $\theta$ varies continuously. On the other hand, if the
combination $R(\theta) + d^2R(\theta)/d \theta^2$ changes sign, then
both $x(\theta)$ and $y(\theta)$ reverse directions with increasing
$\theta$, leading to a cusp at which both segments of the curve meet
tangentially. Such cusps represent the only type of discontinuity that
can appear in the boundary generated by the construction outlined in
Eqs. (\ref{w1})---in the absence of an amputation at a point at which
the boundary crosses itself. If we assume that any cusp generated by
the Wulff construction will make its appearance accompanied by a
swallowtail, then the onset of a cusp is signaled by a change in sign
of $R(\theta) + d^2R(\theta)/d  \theta^2$. This sign change first
appears at the point on the boundary at which the cusp develops.

\section{On variational derivatives}

In the implementation of the generalized version of the Wulff
construction one is, characteristically, faced with the task of taking
the variational derivative with respect to $R(\theta)$ of an integral
of the form
\begin{equation}
\int A\left(\sin\theta,\cos\theta
\right)f\left(x\left(\theta\right), y\left( \theta \right) \right)
\,dS, \label{d1}
\end{equation}
where $A\left(\sin\theta,\cos\theta \right)$ is a polynomial in $\sin
\theta$ and $\cos \theta$, $f\left(x\left(\theta\right), y\left(
\theta \right) \right)$ is a polynomial in $x(\theta)$ and $y(\theta)$
and $dS = \left[R(\theta) + d^2R(\theta)/d \theta^2\right] d\theta$ is
the infinitesimal length element along the boundary. While the
relatively simple dependence of $x$ and $y$ on $\theta$ allows the
variational derivatives to be implemented in a straightforward manner,
the complete and accurate determination of the derivative with respect
to $R(\theta)$ of an integral in which the functions $f(x,y)$ and
$A(\sin \theta, \cos \theta)$ have any but the simplest form becomes
extremely tedious, in the absence of a stratagem that allows for the
evaluation of functional derivatives of non-trivial integrands.
Fortunately, such a stratagem exists.

Consider the following version of the integral above:
\begin{equation}
\int \sin \theta f(x,y)\, dS.  \label{d2}
\end{equation}
First, note that
\begin{eqnarray}
\sin \theta dS &=& \sin \theta
\frac{dS}{d\theta} \nonumber \\ &=&\sin \theta \left[R(\theta) +
\frac{d^2R(\theta)}{d\theta^2}\right] \nonumber \\ &=&
-\frac{dx(\theta)}{d\theta} . \label{d3}
\end{eqnarray}
The integral can thus be rewritten as
\begin{equation}
-\int f\left(x\left(\theta\right),y\left(\theta\right)\right)
\frac{dx}{d\theta}\,d\theta. \label{d4}
\end{equation}
Now, we take the derivative $\delta/\delta R(\theta)$ of the above
integral:
\begin{eqnarray}
\lefteqn{\frac{\delta}{\delta R(\theta)}\int
f\left(x\left(\theta\right),y\left (\theta\right)\right)
\frac{dx}{d\theta}\,d\theta}\nonumber \\ &=& -\int \left\{
\left[\frac{\partial f}{\partial x} \frac{\delta x}{\delta R} +
\frac{\partial f} {\partial y} \frac{\delta y}{\delta
R}\right]\frac{dx}{d\theta} + f(x,y) \frac{d}{d
\theta}\frac{\delta}{\delta R}\right\} d\theta \nonumber \\
&=&-\int\left\{ \left[\frac{\partial f}{\partial x} \frac{\delta
x}{\delta R} + \frac{\partial f} {\partial y} \frac{\delta y}{\delta
R}\right]\frac{dx}{d\theta} - \left[\frac{\partial f}{\partial
x}\frac{dx}{d\theta} + \frac{\partial f}{\partial y}
\frac{dy}{d\theta}\right]\frac{\delta x}{\delta R}\right\} d\theta
\nonumber \\ &=& \int \frac{\partial f}{\partial y}\left[\frac{\delta
x}{\delta R} \frac{dy}{d\theta} - \frac{\delta y}{\delta
R}\frac{dx}{d\theta}\right]d\theta \label{d5}
\end{eqnarray}
Now, given Eqs. (\ref{w1}), we have
\begin{eqnarray}
\frac{\delta x}{\delta R(\theta')}&=&\frac{\delta}{\delta
R(\theta')}\left[R(\theta) \cos \theta - \frac{dR(\theta)}{d\theta}
\sin \theta \right] \nonumber \\ &=&\delta(\theta-\theta') \cos \theta
- \delta'(\theta - \theta') \sin \theta , \label{d6}
\end{eqnarray}
and
\begin{eqnarray}
\frac{\delta y}{\delta R(\theta')} &=& \delta(\theta - \theta') \sin
\theta + \delta' (\theta - \theta') \cos \theta . \label{d7}
\end{eqnarray}
With the use of the above two relations and Eqs. (\ref{w1}) the right
hand side of Eq. (\ref{d5}) becomes
\begin{eqnarray}
\lefteqn{\int \frac{\partial f}{\partial y} \delta(\theta - \theta')
\left[R(\theta)+ \frac{d^2R(\theta)}{d\theta^2}\right]d\theta}
\nonumber \\ &=&\frac{\partial f\left(
x\left(\theta'\right),y\left(\theta'\right)\right)}{\partial
y(\theta')} \left[R(\theta')+ \frac{d^2R(\theta')}{d\theta'^2}\right]
.\label{d8}
\end{eqnarray}
Thus,
\begin{equation}
\frac{\delta}{\delta R(\theta)}\int \sin \theta f(x,y)\,dS =
\frac{\partial f}{\partial y} \left[R(\theta)+
\frac{d^2R(\theta)}{d\theta^2}\right]. \label{d9}
\end{equation}
A similar set of manipulations yields the result
\begin{equation}
\frac{\delta}{\delta R(\theta)}\int \cos \theta
f(x,y)\,dS = \frac{\partial f}{\partial x} \left[R(\theta)+
\frac{d^2R(\theta)}{d\theta^2}\right]. \label{d10}
\end{equation}

Note that the final results hold no matter what form is taken by the
function $f(x,y)$. The method described above can be fruitfully
applied when the integrands are more complicated than the one above.

It is important to note that the steps leading to the final result,
Eq. (\ref{d10}), included integrations by parts, in which the
``perfect derivative'' terms were neglected. If the domain has a
boundary that is free of singularities---especially cusps---one can
easily justify this. However, when cusps are present, so that periodic
boundary conditions on an integral around the circumference of a
domain wall cannot be assumed, more attention to those terms is called
for. The absence of a detailed analysis of the effects of a cusp in
the boundary on the results of an integration by parts is a gap in the
development in this work. Such an analysis is clearly called for.

\begin{figure}
\caption{A drawing of the kind of cusped domains seen in experiments.
Such domains contain liquid condensed regions in an environment of the
liquid expanded phase of a Langmuir monolayer. Figure 1a depicts the
domain as it is seen in fluorescence microscopy. Figure 1b adds the
``virtual boojum'' texture that is conjectured to be responsible for
the cusp.}
\end{figure}
\begin{figure}
\caption{The various virtual singularity textures that emerge from
exact solutions to Eqs. (2.2) and (2.3) when only one of the $a_n$'s
in the surface energy in non-zero and the domain is circular.  Figure
2a: $n=1$, 2b: $n=2$, 2c: $n=3$, 2d: $n=4$. The singularities sit
outside the domain, except in the limit of inifinite $\kappa$. For
ease of depiction, they are shown on the domain boundary.}
\end{figure}
\begin{figure}
\caption{The quantities $\theta$ and $R(\theta)$, as defined in Eqs.
(3.1).} \end{figure} \begin{figure} \caption{The shape of a domain,
when the surface tension has the anisotropic dependence on $\theta$,
$e^{\beta \cos(\theta)}$, and $x(\theta)$ and $y(\theta)$ are given by
Eqs. (3.12). As indicated in the Figure, the three cases depicted are
$\beta=0.5$, $\beta=1$ and $\beta=1.5$.}
\end{figure}
\begin{figure}
\caption{The function ${\rm Re}\left[{\rm{\bf X}}(\psi)\right]$, as
defined in Eq.(6.21)} \end{figure} \begin{figure} \caption{A plot of
the excluded angle $\Delta \psi = \pi - \psi_{{\rm in}}$ versus the
radius of the domain. The radius of the domain is expressed in units
of $\kappa/a_1$. The ratio $a_2/\sigma_0$ has been set equal to 0.05.
The quantities are defined in Section VI. The quantity $\psi_{{\rm
in}}$ is the interior angle of the cusp, and is shown in Figure 7}
\end{figure}
\begin{figure}
\caption{The domain when $a_2/\sigma_0=0.25$ and $R_0a_1/\kappa=5$.
For definitions of the quantities, see Section VI. The interior angle,
$\psi_{{\rm in}}$, is indicated in the Figure.}
\end{figure}
\begin{figure}
\caption{A comparison between the curve for the excluded cusp angle
displayed in Figure 6 and data obtained by Schwartz, Tsao and Knobler
(see ref. 12). The axes are both logarithmic and scales are adjusted
to obtain a by-eye fit. No attempt was made to optimize agreement by
adjusting energy parameters in the theoretical result}
\end{figure}

\end{document}